# Discovery of a Low-redshift Hot Dust-obscured Galaxy

Guodong Li[1,3], Chao-Wei Tsai[1,2,3], Daniel Stern[4], Jingwen Wu[1,3], Roberto J. Assef[5], Andrew W. Blain[6],
Tanio Díaz-Santos[7,8], Peter R. M. Eisenhardt[4], Roger L. Griffith[4], Thomas H. Jarrett[9], Hyunsung D. Jun[10],
Sean E. Lake[1], and M. Lynne Saade[11]

[1] National Astronomical Observatories, Chinese Academy of Sciences, 20A Datun Road, Beijing 100101, People's Republic of China; cwtsai@nao.cas.cn, jingwen@nao.cas.cn
[2] Institute for Frontiers in Astronomy and Astrophysics, Beijing Normal University, Beijing 102206, People's Republic of China
[3] University of Chinese Academy of Sciences, Beijing 100049, People's Republic of China
[4] Jet Propulsion Laboratory, California Institute of Technology, 4800 Oak Grove Drive, Pasadena, CA 91109, USA
[5] Instituto de Estudios Astrofísicos, Facultad de Ingeniería y Ciencias, Universidad Diego Portales, Av. Ejército Libertador 441, Santiago, Chile
[6] Department of Physics and Astronomy, University of Leicester, LE1 7RH Leicester, UK
[7] Institute of Astrophysics, Foundation for Research and Technology-Hellas (FORTH), Heraklion, GR-70013, Greece
[8] School of Sciences, European University Cyprus, Diogenes Street, Engomi, 1516 Nicosia, Cyprus
[9] Department of Astronomy, University of Cape Town, Private Bag X3, Rondebosch, 7701, South Africa
[10] SNU Astronomy Research Center, Astronomy Program, Department of Physics and Astronomy, Seoul National University, Seoul 08826, Republic of Korea
[11] Department of Physics and Astronomy, University of California at Los Angeles, Los Angeles, CA 90095-1547, USA
Received 2022 December 29; revised 2023 June 26; accepted 2023 June 26; published 2023 November 21

## Abstract

We report the discovery of the hyperluminous, highly obscured active galactic nuclei (AGN) WISE J190445.04 +485308.9 (W1904+4853, hereafter, $L_{\rm bol} \sim 1.1 \times 10^{13} L_\odot$) at $z = 0.415$. Its well-sampled spectral energy distribution (SED) is dominated by infrared dust emission, though broad emission lines are detected in the optical spectra. These features suggest that W1904+4853 contains an actively accreting supermassive black hole hidden in its dusty cocoon, resembling the observed properties of hot dust-obscured galaxies (Hot DOGs), a population previously only identified at $z > 1.0$. Using the broad component of the Mg II 2798 Å emission line, we estimate a black hole mass of $\log(M_{\rm BH}/M_\odot) = 8.4 \pm 0.4$. The corresponding Eddington ratio ($\eta$) of $1.4^{+1.3}_{-0.7}$ implies that the central black hole accretion is at the theoretical limit of isotropic accretion. The rest-frame UV-optical SED also indicates that the host galaxy of W1904+4853 harbors strong star formation activity at the rate of $6-84\,M_\odot\,{\rm yr}^{-1}$ with an independent estimate of star formation rate up to $\sim 45\,M_\odot\,{\rm yr}^{-1}$ using the [O II] emission line. With an estimated stellar mass of $3 \times 10^{10}\,M_\odot$, the host galaxy appears to be a starburst system with respect to the main sequence of the star-forming galaxies at the same redshift. Although blueshifted and asymmetric [O III] emission provides evidence of an outflow, we estimate it to be an order of magnitude smaller than the star formation rate, indicating that the current obscured AGN activity at the center has not yet produced significant feedback on the host galaxy star formation activity. W1904+4853 supports the interpretation that Hot DOGs are a rare transitional phase of AGN accretion in galaxy evolution, a phase that can persist into the present-day Universe.

*Unified Astronomy Thesaurus concepts:* Ultraluminous infrared galaxies (1735); Supermassive black holes (1663); Active galaxies (17)

## 1. Introduction

The framework of massive galaxy evolution through major mergers has been used to explain the coevolution of supermassive black holes (SMBHs) and their host galaxies (Sanders et al. 1988; Barnes & Hernquist 1992; Schweizer 1998; Hopkins & Beacom 2006; Jogee 2006; Hopkins et al. 2008). In that scenario (e.g., Hopkins et al. 2008), tidal torques caused by major mergers induce gas infall toward the center of the galaxy, initiating a central starburst and triggering rapid growth of the SMBH. Intense radiative feedback from the SMBH accretion suppresses gas falling toward the central region, driving gas and dust out of the surrounding area, and shutting down nearby star formation in the host galaxy. During this short *blowout* phase, the actively accreting SMBH is highly obscured and very luminous. Therefore, this phase is likely associated with highly dust-reddened and IR-luminous quasars.

Surveying the entire sky at mid-infrared (MIR) wavelengths, the Wide-field Infrared Survey Explorer (WISE) aimed to identify the most luminous dusty galaxies in the Universe as one of its main scientific objectives (Wright et al. 2010). Luminous dusty systems are likely powered by obscured active galactic nuclei (AGN). In such galaxies, the dust-wrapped accreting SMBH heats the surrounding dust to high temperatures, releasing most of the radiation at IR wavelengths. By selecting systems with robust detections in the WISE W3 and W4 bands (12 and 22 $\mu$m, respectively) but marginal or no detections in the W1 and W2 bands (3.4 and 4.6 $\mu$m, respectively), a population of hot dust-obscured galaxies (Hot DOGs; Eisenhardt et al. 2012; Wu et al. 2012) powered by highly obscured AGN was discovered in the past decade. Compared with dust-obscured galaxies (DOGs; Dey et al. 2008), Hot DOGs are significantly redder in WISE passbands due to their hotter dust emission heated by the central AGN (Wu et al. 2012; Bridge et al. 2013), and have much higher bolometric luminosities (Tsai et al. 2015).

With their number density peaking at $z \sim 2-3$ (Eisenhardt et al. 2012; Assef et al. 2015; Tsai et al. 2015), Hot DOGs have intrinsic total luminosities generally above $10^{13} L_\odot$, with ~10% exceeding $10^{14} L_\odot$ (Tsai et al. 2015). The Hot DOG WISE J2246-0526 at $z = 4.601$ (Tsai et al. 2015; Díaz-Santos et al. 2018) is the most luminous galaxy currently known. Follow-up studies show that most Hot DOGs have similar spectral energy distributions (SEDs; Wu et al. 2012; Tsai et al. 2015), characterized by a broad MIR plateau and a falloff at far-infrared (FIR) to submillimeter wavelengths. The high MIR luminosity of Hot DOGs implies a substantial contribution from much hotter dust than seen in other IR-luminous galaxies. Combined with the evidence of observed broad Balmer lines and high ionization lines, the existence of hot dust suggests that their extreme luminosities are powered by AGNs. The AGNs are highly obscured ($A_v \sim 20$–60 mag; Assef et al. 2015), often Compton thick in the X-ray (Stern et al. 2014; Piconcelli et al. 2015; Assef et al. 2016). SMBH mass measurements from the broad emission line widths further indicate that their AGN accretion is around the Eddington limit (Tsai et al. 2018; Wu et al. 2018). Hot DOGs show spectral signatures of ionized gas outflow (Finnerty et al. 2020; Jun et al. 2020), suggesting that Hot DOGs might be in the *blowout* phase of galaxy evolution before the emergence of visible quasars. This transitional phase coincides with the most rapid SMBH growth and AGN feedback, and hence is important for understanding the coevolution of SMBHs and their hosts.

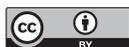

High luminosities driven by high accretion rates in Hot DOGs (Tsai et al. 2018; Wu et al. 2018) are also seen in $z \sim 6$ quasars (e.g., De Rosa et al. 2011). These two types of systems are the most luminous objects with the maximal accretion rates in their cosmic epochs. This high accretion rate stage might also exist at lower redshift, such as in the case of W1036+0449, a $z = 1.0$ Hot DOG (Ricci et al. 2017). However, it is yet unclear whether Hot DOGs exist in the more modern Universe and whether they are experiencing similarly rapid SMBH growth while becoming hyperluminous. In light of the comparable number density of high-$z$ Hot DOGs to that of hyperluminous type 1 AGN at redshift $2 < z < 4$ (Stern et al. 2014; Assef et al. 2015), a naive assumption can be made that the number density of low-$z$ systems is also similar to type 1 AGN at equivalently high luminosity levels. If so, the estimated number of low-$z$ Hot DOGs would be low, making it challenging to identify them. However, finding such rare low-$z$ systems would allow close-up investigations into the composition of gas and dust in these maximally accreting systems with strong feedback at much higher resolution and greater detail, shedding light on galaxy evolution during the most violent evolutionary phases.

This paper reports the discovery of the dust-obscured quasar WISE J190445.04+485308.9 (W1904+4853, hereafter) at $z = 0.415$. The optical spectrum, together with a description of other observations of W1904+4853 used in this paper, are presented in Section 2. Section 3 presents the analysis of the SED and spectroscopy. Discussion of the black hole (BH) mass, ionized gas outflow, Eddington ratio, and host galaxy are in Section 4. Section 5 summarizes our work. In this paper, we adopt a cosmology with $H_0 = 70$ km s$^{-1}$ Mpc$^{-1}$, $\Omega_m = 0.3$, and $\Omega_\Lambda = 0.7$.

## 2. Observations and Data Reduction

Inspired by the discovery of low-metallicity blue compact dwarf (BCD) galaxies via their unusually red WISE colors (Griffith et al. 2011), a search for additional BCDs was conducted using the selection criteria of $1.9 < W1\text{–}W2 < 3.0$, $2.9 < W2\text{–}W3 < 5.3$, and optical $R < 19.5$ mag. Few objects are found in this region of WISE color space (see Figure 1 in Eisenhardt et al. 2012).

W1904+4853 was identified by this selection, with its unusual colors of W1–W2 = $2.39 \pm 0.04$ and W2–W3 = $5.06 \pm 0.03$, and its $R$-band magnitude of 19.4 in the SuperCOSMOS Science Archive (SSA; Hambly et al. 2001).

### 2.1. Spectroscopic Observations

An optical spectrum for W1904+4853 was first obtained with the Low-Resolution Imaging Spectrometer (LRIS; Oke et al. 1995) at the Keck I Telescope on UT 2010 November 9. The spectrum was obtained with a total of 250 s integration on source with a 1″5 slit (seeing ~1″5). The spectrum only covers 5800–10,320 Å due to the temporary malfunction of the blue arm camera of the LRIS during the observation. Strong H$\alpha$, H$\beta$, H$\gamma$, and [O III] 4959 and 5007 Å lines are well detected at redshift $z = 0.415$.

To investigate the Mg II line, we reobserved W1904+4853 with the Double Spectrograph (DBSP) on the Palomar 200 inch Hale Telescope on UT 2020 July 24. We limit the spectrum from the blue and red arms of the spectrograph to rest-frame wavelengths 2500–3800 Å and 4100–7000 Å respectively, to minimize sensitivity losses at the edges of the observed spectral ranges. The 1″0 slit set to align with the parallactic angle of 110° was placed on the target, centered on the position R.A. =19$^h$04$^m$45$^s$.04, decl. =+48°53′08″.9 (J2000). The full DBSP spectrum with a total of 1200 s integration at a mean air mass of 1.1. The reduced spectrum is shown in Figure 1. The spectrum was processed and analyzed with IRAF, following the standard reduction procedures to account for bias and flat fielding. The spectrum was corrected for telluric absorption features and flux calibrated using the standard star BD+28 4211. The flux difference between the LRIS spectrum and the DBSP spectrum is less than 10%. The detailed line measurement process and redshift determination are discussed in Section 3.4. We note that the 2D spectrum of W1904+4853 from Palomar/DBSP does show a marginal extension along the slit. The FWHMs of emission lines are generally marginally extended at 1″6 and the continuum is slightly bigger with a size of 1″9, under the seeing condition of 1″5.

### 2.2. Multiband Photometry

To construct the SED of W1904+4853 and estimate its bolometric luminosity, we gathered its photometric data at UV-to-submillimeter wavelengths from the GALEX 10.17909/ths3-nr82, Pan-STARRS 10.17909/s0zg-jx37, UKIRT, WISE 10.26131/IRSA1, AKARI 10.26131/IRSA180, IRAS 10.26131/IRSA11, and Spitzer 10.17909/mkcw-8w83 surveys. We also obtained near-IR (NIR) photometry from United Kingdom Infrared Telescope (UKIRT) and submillimeter flux measurements from James Clerk Maxwell Telescope (JCMT). The summary of the multiwavelength photometric data are given in Table 1 and plotted in Figure 2.

#### 2.2.1. NIR and Submillimeter Observations

We imaged W1904+4853 with UKIRT/Wide Field Camera (WFCAM) in the $H$ and $K$ filters on UT 2020 July 29 under photometric conditions (seeing ≃0″75). For each filter, we used a five-jitter dither pattern with $2 \times 2$ microstepping to

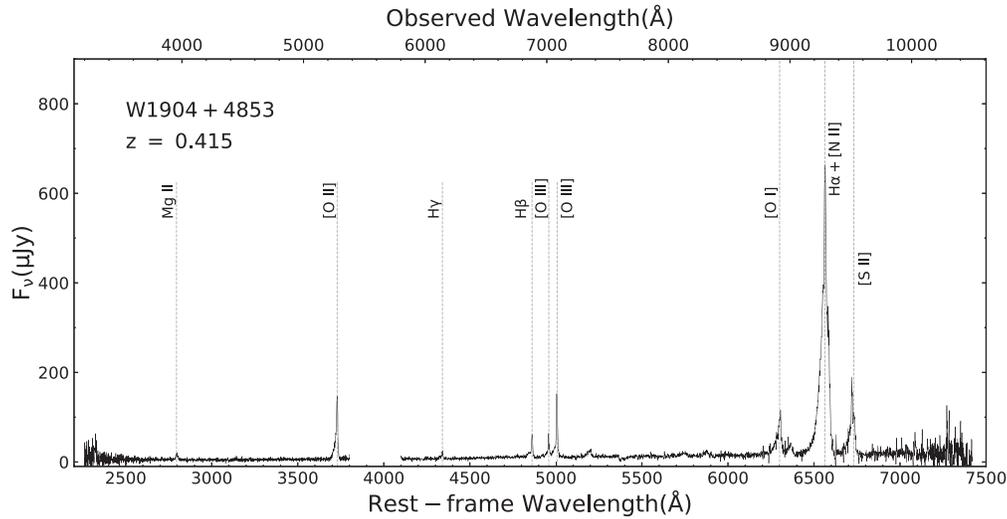

**Figure 1.** Optical spectrum of W1904+4853 obtained at Palomar Observatory. The spectrum covers wavelengths 3200–10,500 Å in the observed frame, with a gap from 5500–5700 Å due to the dichroic mirror. The emission lines labeled on the plot are for the spectroscopic redshift of $z = 0.415$, corresponding to rest-frame wavelength coverage of 2260–7420 Å.

**Table 1**
Flux Densities of W1904+4853

| Band | Effective Wavelength ($\mu$m) | Flux Density ($\mu$Jy) | Corr Flux ($\mu$Jy) |
|---|---|---|---|
| GALEX.nuv | 0.23 | $8 \pm 1$ | $6.7 \pm 1.6$[a] |
| INT.WFC.U | 0.36 | $7.3 \pm 0.6$ | ... |
| INT.WFC.H$\alpha$ | 0.66 | $32 \pm 2$ | ... |
| PS1.g | 0.48 | $22.2 \pm 0.7$ | $16 \pm 1$ |
| PS1.r | 0.62 | $32.0 \pm 0.6$ | $29.5 \pm 0.7$ |
| PS1.i | 0.75 | $47.4 \pm 0.7$ | $38 \pm 1$ |
| PS1.z | 0.87 | $70 \pm 2$ | $42 \pm 3$ |
| PS1.y | 0.96 | $214 \pm 10$ | $49 \pm 16$ |
| UKIRT.J | 1.25 | $83 \pm 12$ | ... |
| UKIRT.H | 1.65 | $133 \pm 7$ | ... |
| UKIRT.K | 2.20 | $209 \pm 9$ | ... |
| Spitzer.IRAC.I1 | 3.54 | $700 \pm 40$ | ... |
| Spitzer.IRAC.I2 | 4.48 | $2.4 \pm 0.2 \times 10^3$ | ... |
| WISE.W1 | 3.35 | $547 \pm 13$ | ... |
| WISE.W2 | 4.60 | $2.72 \pm 0.05 \times 10^3$ | ... |
| WISE.W3 | 11.56 | $54.4 \pm 0.8 \times 10^3$ | ... |
| WISE.W4 | 22.09 | $157 \pm 3 \times 10^3$ | ... |
| Akari.WIDE-S | 76.90 | $580 \pm 170 \times 10^3$ | ... |
| IRAS.60 | 51.99 | $560 \pm 30 \times 10^3$ | ... |
| IRAS.100 | 95.30 | $750 \pm 180 \times 10^3$ | ... |
| JCMT.450 | 448 | $< 73 \times 10^3$ | ... |
| JCMT.850 | 858 | $6.3 \pm 2.9 \times 10^3$ | ... |
| VLA.S | 74900 | $9.7 \pm 0.2 \times 10^3$ | ... |

**Note.** "Flux Density" represents the observed value. "Corr Flux" is the corrected flux, removing the contribution of emission lines using the observed spectra except for the case of GALEX NUV in which estimated equivalent widths of two emission lines are used.

[a] The flux correction for GALEX NUV uses average equivalent widths of C III] and C IV emission lines from high-$z$ Hot DOGs. The uncertainties are derived from the standard deviations of observed equivalent widths (P. R. M. Eisenhardt et al. 2023, in preparation).

produce the final stacked image from individual frames with 100 s integration times. We use the reduced and calibrated photometric data with a 4″ diameter aperture from the WFCAM Science Archive. The flux encompassed within a 0.″8 (∼4 kpc) aperture accounts for <70% of the total energy for W1904 +4853 in both H- and K-band images, showing a marginal spatial extension. However, the signal-to-noise ratio (S/N) is inadequate for making definitive morphological claims.

To constrain the FIR SED, we obtained a submillimeter observation of W1904+4853 with JCMT/Submillimeter Common-User Bolometer Array 2 (SCUBA-2; Holland et al. 2013) on UT 2021 February 22. The standard constant velocity Daisy scanning pattern was used to maximize sensitivity on the selected observing field of our unresolved target. With a 2 hr on-source integration in band-3 weather conditions (1.58 mm < precipitable water vapor < 2.58 mm), W1904+4853 shows a marginal detection ($6.3 \pm 2.9$ mJy) at 850 $\mu$m and an upper limit at 450 $\mu$m (rms ∼73 mJy beam$^{-1}$).

### 2.2.2. Archival Photometric Data

The UV photometry data were obtained from the GALEX Medium-deep Sky Catalog in General Release 6 (Seibert et al. 2012). The FUV data were excluded in our analysis due to artifacts indicated by the artifact flags and extraction flags (Bianchi et al. 2017). We adopt NUV = $21.60 \pm 0.14$ (AB).

The optical U-band photometry and H$\alpha$-band data of W1904 +4853 are from the public science archive of the Kepler-INT Survey (Data Release 2; Greiss et al. 2012), and the five broadband (g, r, i, z, and y) data are from the Pan-STARRS Survey (Data Release 1; Chambers et al. 2016).

We used the WISE MIR data from the AllWISE Release database (Cutri et al. 2013). For the FIR photometry, we adopted the fluxes at 60 and 100 $\mu$m from IRAS. Akari reports a detection of W1904+4853 at 63 $\mu$m albeit with an uncertainty reported as "NULL," implying its flux density is close to the detection limit (Murakami et al. 2007; Yamamura et al. 2010). Thus, we did not include this Akari photometric data in the SED analysis. W1904+4853 has also been imaged by the Spitzer Space Telescope at 3.6 and 4.5 $\mu$m as part of the Spitzer Kepler Survey (SpiKeS; Werner et al. 2013), displaying an unresolved and pointlike structure. We conducted photometric measurements for W1904+4853 with an 8″ diameter aperture using the reduced and calibrated images obtained from the IPAC/Spitzer Science Archive. The S-band flux of W1904 +4853 is obtained from the VLA Sky Survey (VLASS; Lacy

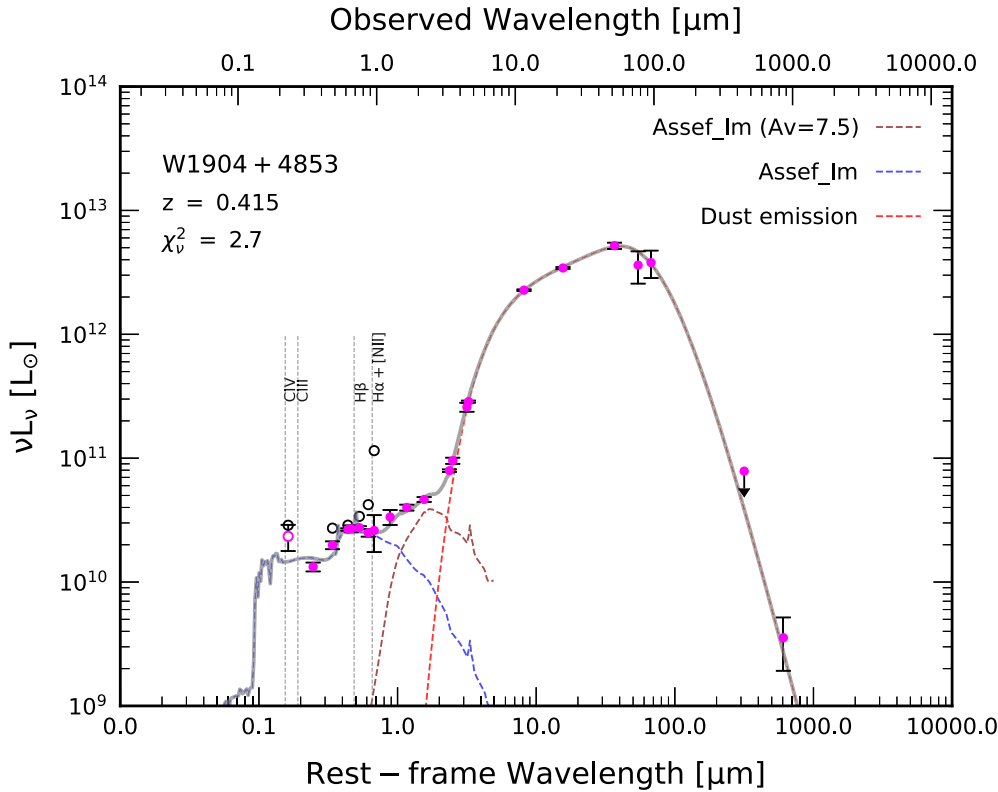

**Figure 2.** Best-fit SED template model (heavy gray line) of W1904+4853. The photometric data points are extinction corrected and emission line subtracted. The black open circles represent the photometry before the emission-line corrections. See Section 2.2 for details. The magenta open circle with an error bar is the line-subtracted GALEX NUV flux, assuming it is also affected by strong emission lines. See Section 3.1 for details. The SED modeling uses templates of irregular galaxies with $A_V = 7.5^{+1.1}_{-1.0}$ (brown) and starburst components (blue) from Assef et al. (2010), and the single power-law (SPL) dust distribution model (red) from C.-W. Tsai et al. (2023, in preparation) with $p = -0.82 \pm 0.02$, $T_h = 539^{+6}_{-6}$ K, and $T_c = 37^{+3}_{-3}$ K. The parameters of SPL ($p$, $T_h$, and $T_c$) represent the power-law index for dust grain number density with radius, and the upper and lower limits of the dust temperature (see Section 3.2).

et al. 2020; Gordon et al. 2021). The results are listed in Table 1.

### 3. Analysis

#### 3.1. Photometry Corrected for Strong Emission Lines

W1904+4853 has strong and broad emission lines and faint continuum emission at optical wavelengths, similar to many high-$z$ Hot DOGs (Wu et al. 2012; Assef et al. 2015; Tsai et al. 2015). These observed strong lines significantly contribute to the optical photometry and can obstruct the SED fitting if the used templates are constructed with weak-line conditions. To better model the SED of its host galaxy, we estimate the line-free continuum photometry by subtracting the emission-line contribution to the broadband photometry. The line-subtracted continuum flux and uncertainties are listed in Table 1.

We first assume that the broad emission lines are from the centrally concentrated and unresolved core region under the seeing condition, and the stellar continuum emission of the host galaxy is extended and marginally resolved as seen in the Pan-STARRS images (deconvolved FWHM ∼1″.5 in the r band). The difference in the slit loss values for the components with different spatial extents in the DBSP's 1″.0-slit spectrum and the seeing of ∼1″.5 are significant and used to derive the intrinsic line-to-continuum ratios. These intrinsic ratios, which are lower than in the observed spectrum, are combined with the filter bandpasses (see Figure 3) to estimate the line-free continuum photometry. The Hα filter is not significantly contaminated by emission lines and therefore does not need this correction. While Mg II falls in the U band, it only contributes ∼2% to the photometry, which is small compared to the ∼8% uncertainty of the U-band flux, so we did not correct this band.

We consider the possibility of the NUV band being affected by strong emission lines. The UV fluxes in the GALEX NUV band is significantly higher than the best-fit model shown in Figure 2 (black open circle). One possibility is that the elevated NUV flux may be contaminated by the strong and redshifted C IV 1549 Å and C III] 1909 Å lines at the range of 1400–2800 Å, which are often seen in high-$z$ Hot DOGs ($z > 1.5$; Wu et al. 2012). We estimated C III] and C IV emission contributes 16% of the observed NUV flux, based on the averaged equivalent widths of these lines seen in high-$z$ Hot DOGs (P. R. M. Eisenhardt et al. 2023, in preparation), or up to 45% using the maximum equivalent widths seen.

#### 3.2. SED Fitting

The red MIR color, strong broad emission lines, and high luminosity of W1904+4853 suggest this system is mainly powered by an obscured AGN, as is thought to be true for high-redshift Hot DOGs (Eisenhardt et al. 2012; Assef et al. 2015; Tsai et al. 2015). Except for a few systems identified as *blue Hot DOGs* (Assef et al. 2016, 2020, 2022)—Hot DOGs with blue excess in their SEDs due to a small fraction of AGN blue light scattered through the dust cocoon, the emission of the powerful AGN is almost fully obscured. The rest-frame optical SED of Hot DOGs generally represents the host galaxy properties with little contribution from the deeply buried AGN.

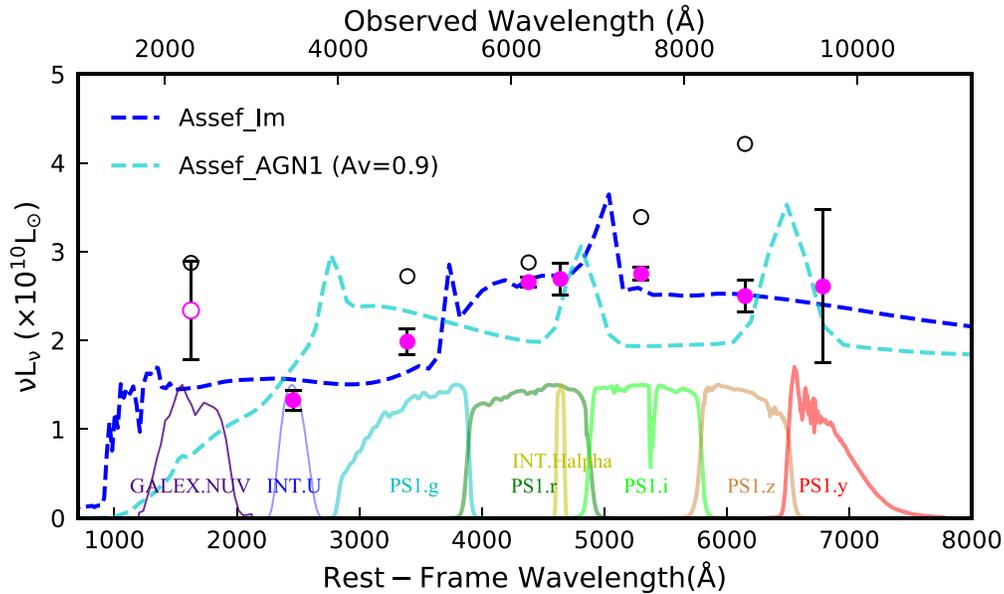

**Figure 3.** Zoomed-in UV-optical SED of W1904+4853. As in Figure 2, the filled magenta data points denote the emission line subtracted, while black open circles represent the photometry before the emission-line corrections. Strong H α emission in the y band puts the observed photometry off-scale, so only the emission-line-subtracted y band is shown here. Filter transmissions for the photometric points are displayed at the bottom of the panel. The blue dashed line shows the unobscured starburst (Im) component from the case 1 best-fit SED model (also shown in Figure 2), which dominates the UV optical. The cyan dashed line shows the mildly obscured ($A_V \sim 0.9$ mag) AGN (Assef et al. 2010) component from the case 3 model, which again dominates the UV optical, but fails to reproduce the shape of the continuum.

To decompose the stellar components in the host galaxy, we employed the approach described by Assef et al. (2008, 2010) in which the SED of a galaxy is a linear combination of empirical spectral templates (Assef et al. 2010). The four empirical SED templates—"AGN" for an unreddened AGN, "E" for an elliptical galaxy (old stellar population), "Sbc" for a spiral galaxy (intermediate-aged stellar population), and "Im" (young starburst population) for an irregular galaxy—are used to model the SED of an AGN and its host galaxy in the rest-frame wavelength range from 0.03–30 μm (Assef et al. 2010). Each template can have an independent reddening applied.

These templates do not cover wavelengths longer than 30 μm in the rest frame. In addition to the host galaxy components, the SED of W1904+4853 (see Figure 2), like that of the Hot DOG population, is dominated by hot dust beyond a few microns in the rest frame, with significant emission to beyond 100 μm (Tsai et al. 2015). Therefore, to model the IR SED (>2 μm) we add an SPL dust distribution component, which was developed to describe dust emission for Hot DOGs (C.-W. Tsai et al. 2023, in preparation). The dust emission model assumes that dust number density is a simple power-law function of radius from the galaxy center, and the thermal dust emission is in equilibrium with the radiation received by the dust grain. In addition to the slope of the power law, the SPL model uses upper and lower dust temperature limits to represent the inner and outer cutoff of the dust distribution. The detailed best-fit SPL model is described in Section 4.1.

The dust component dominates the bolometric luminosity with its MIR and far-IR (FIR) emission. Other characteristics in the emission-line corrected SED of W1904+4853 are a steeper ascent in the rest frame NIR than expected for an unobscured stellar population, and a UV-optical continuum suggestive of a stellar population with relatively little obscuration. Motivated by these features, we modeled the SED for three different cases to assess the roles of AGN, dust, and star formation in W1904+4853. Each case assumed three components were present, one of which was the SPL dust model, with reddened Assef et al. (2010) templates used for the other two components. We assumed $R_V = 3.1$ and an SMC reddening law (Gordon et al. 2003) at short wavelengths and a Milky Way reddening law (Fitzpatrick & Massa 2007) at longer wavelengths.

Prior to the SED modeling process, the flux of each band was corrected for extinction from dust in the Milky Way ($A_V = 0.19$ mag; Schlegel et al. 1998; Fitzpatrick & Massa 2007). We fit the SED using the Markov Chain Monte Carlo (MCMC) method provided by the Python package emcee (Foreman-Mackey et al. 2013), solving for the amplitude of the two template components and the SPL component simultaneously, along with the obscuration (for case 3 only the AGN obscuration was varied). A simple Gaussian function was used for the likelihood function with uninformative priors. The median value of the MCMC samples was adopted as the best-fit result of each parameter and the quoted "1σ" uncertainties are based on the samples' 16th and 84th percentiles in the marginalized distributions.

Case 1 assumes, in addition to the SPL hot dust component with the power-law index of dust radius distribution $p = -0.82 \pm 0.02$, and higher temperature limit $T_h = 539^{+6}_{-6}$ K, and lower temperature limit $T_c = 37^{+3}_{-3}$ K, host galaxy emission dominated by a young starburst population partially obscured by dust. We used two Im models to represent the young stellar population, with extinction applied to only one of them. This model as shown in Figure 2, provided the best SED fit ($\chi^2_\nu = 2.7$) of the three cases considered. The obscured starburst population has a substantial extinction of $A_V \sim 7.5$ mag from the SED fit. The extinction toward a fraction of the young starburst population (Im template) is more reasonable than for the E or Sbc templates. This obscured stellar component provides a good match to the observed NIR bands, while the unobscured starburst dominates the shorter wavelength bands. In this case, the AGN is completely obscured and contributes no emission in the UV-optical wavelengths.

Case 2, like case 1, also assumes the SPL hot dust component and an unobscured starburst population are present, but replaces the obscured starburst with light transmitted through the dust around an obscured AGN. This model intends to use an obscured AGN SED to explain the relatively elevated NIR luminosity. The best-fit result again has a starburst population dominating the SED between 0.1 and 1.0 $\mu$m. The dusty AGN with extinction $A_V \sim 2.5$ mag dominates the observed $J$, $H$, and $K$ bands, but the fit here is not as good as for case 1, resulting in a significantly worse reduced chi-squared ($\chi^2_\nu = 10.5$).

Case 3 again assumes the SPL component is present, as well as a starburst component with a given $A_V \sim 7.5$ mag obscuration, but replaces the unobscured starburst with a lightly obscured AGN component. This model mainly attributes the optical and UV emission from a lightly obscured AGN. The amplitude of the obscured starburst is a free parameter in the fitting. Some escaping AGN light is motivated by the broad emission lines seen in the optical, but the best-fit obscuration ($A_V \sim 0.9$ mag) for an AGN producing the short wavelength light has a much higher reduced chi-squared ($\chi^2_\nu = 34$) than other cases. Like case 2, the fit in the observed NIR is worse than for case 1, and as shown in Figure 3, a mildly obscured AGN system cannot explain the change in the continuum from rest UV-to-optical wavelengths. However, the UV-optical continuum can be modeled by a young stellar population.

In summary, we examined three scenarios in the SED-fitting analysis. Case 1 includes an AGNs dust emission SPL model, an unobscured starburst model, and an obscured starburst population ($A_V \sim 7.5$ mag) that dominates NIR emission. Case 2 contains an AGNs dust emission SPL model, an unobscured starburst model, and a highly obscured AGN ($A_V \sim 2.5$ mag) that dominates NIR photometry. Case 3 has an AGNs dust emission SPL model, a lightly obscured AGN ($A_V \sim 0.9$ mag) dominating optical-NIR photometry, and an obscured starburst ($A_V \sim 7.5$ mag) contributing mainly to NIR wavelengths. The combination of SED models in case 1 yields the best-fit results. We adopt that scenario of case 1 in which an extremely obscured AGN with only hot dust emission is in a host galaxy with both obscured and unobscured starburst stellar population in the discussions in the rest of the paper. We cannot rule out, however, that light from the AGN scattered by gas or dust may contribute to the UV-to-NIR fluxes.

### 3.3. Luminosity Estimates and Variability

The bolometric luminosity of W1904+4853 was calculated using the emission-line subtracted photometric data shown in Figure 2. We obtain a lower limit of $L_{bol} = 1.0 \times 10^{13}~L_\odot$ for the bolometric luminosity using a power-law interpolation between photometric data, assuming no contributions from emission at wavelengths shorter or longer than the observed data. If the best-fit SED model (heavy gray line in Figure 2) between wavelengths of 0.03 and 1000 $\mu$m in the rest frame is adopted, the estimated bolometric luminosity is $1.1 \times 10^{13}~L_\odot$, which is considered more representative of the total luminosity of W1904+4853 and is adopted as the bolometric luminosity in this paper. The IR dust emission, $L(2–1000~\mu m)$, dominates this total luminosity (>99%). The systemic uncertainty of the total luminosity is considered to be $\sim 20\%$ (Tsai et al. 2015) from the assumption of a slow-varying SED, which is more significant than the statistical uncertainty (<10%). We note that the $S$-band (2–4 GHz) luminosity of $\sim 6 \times 10^7~L_\odot$ for W1904+4853 does not significantly contribute to the bolometric luminosity. In comparison with the optical luminosity ($>10^{10}~L_\odot$), W1904+4853 can be classified as a radio-quiet system.

To check if W1904+4853 is variable in the MIR, we investigated its data in the unTimely catalog (Meisner et al. 2023), which contains 16 epochs of time-resolved WISE and NEOWISE co-adds from 2010–2020. We found that the difference in W2 magnitude between the first and last unTimely epochs is $0.077 \pm 0.012$, which is much higher than sources with comparable brightness in the same region of sky. Considering all 646 sources with $11.5 < W2$ (first epoch) $<12.5$ in the same $1°.56 \times 1°.56$ WISE tile, only 10 sources showed a larger W2 change. After the second epoch, corresponding to 2010 November, the source has remained stable to $\pm 0.02$ mag in W2. The difference between the first and last epochs in W1 magnitude is $0.000 \pm 0.022$, which is typical for sources of comparable W1 flux in the same WISE tile. WISE shows this source faded modestly in W2 ($< 10\%$) in 2010, after which the source has remained stable. This result of a greater (but still small) variation in W2 compared to W1 is consistent with our SED modeling (Section 3.2), which AGN emission strongly dominates W2, while the host galaxy contributes significantly to W1. This suggests that, besides emission-line contamination, the bright GALEX photometry might also be due to a brighter episode in the source's past. We also examined the optical photometric data from Kepler-INT survey ($g$, $r$, and $i$; Greiss et al. 2012) and DESI Legacy Imaging Surveys ($g$, $r$, and $z$; Dey et al. 2019), and find no significant ($>4\sigma$) optical variability. We note that the $z$-band flux in the Legacy survey is significantly higher than that in the Pan-STARRS survey because the strong H$\alpha$ line of W1904+4853 fully falls into the range of a wider $z$ band in the Legacy survey.

### 3.4. Emission Line Fitting

The optical spectrum of W1904+4853 (Figure 1) shows well-detected Mg II, [O II], H$\gamma$, H$\beta$, [O III]4959 Å, [O III] 5007 Å, [O I], H$\alpha$, [N II], and [S II] emission lines. We used a power law to model the continuum emission, and single Gaussian, or two Gaussian profiles to model the emission lines. Table 2 summarizes the broad/narrow fitting schemes used for the selected lines. The emission-line fitting results are shown in Figure 4 and summarized in Table 3.

The forbidden lines of W1904+4853 are generally broadened, asymmetric, and blueshifted. The broadened forbidden lines are generally considered evidence of ionized gas outflowing from an AGN at $\gtrsim 10^3$ km s$^{-1}$ (Veilleux et al. 1994; Crenshaw & Kraemer 2000) instead of the virialized motions of clouds in the broad-line region (BLR). Following the analysis presented in Jun et al. (2020) for modeling the spectra of 12 Hot DOGs, the Balmer emission lines (H$\alpha$, H$\beta$, H$\gamma$) and [O I], [O II], [O III], as well as [S II] doublets of W1904+4853 are modeled with a single narrow (FWHM $\leqslant$ 1200 km s$^{-1}$) and a single broad (1200 km s$^{-1}$ $\leqslant$ FWHM $\leqslant$ 10,000 km s$^{-1}$) Gaussian profile. For [O I], [O III], [S II], and

**Table 2**
Fitting Schemes for Emission Lines

| Line | Gaussians | Fixed Line Spacing | FWHM | Flux Ratio |
|---|---|---|---|---|
| Mg II λ2798 | B+Fe II | ⋯ | ⋯ | ⋯ |
| [O II] λ3727 | N+B | ⋯ | ⋯ | ⋯ |
| Hγ λ4340 | N+B | ⋯ | ⋯ | ⋯ |
| Hβ λ4861 | N+B | ⋯ | ⋯ | ⋯ |
| [O III] λ4959/5007 | N+B | Yes | Fixed | 2.99 |
| [O I] λ6300/6363 | N+B | Yes | Fixed | ⋯ |
| [N II] λ6548/6583 | N | Yes | Fixed | 2.96 |
| Hα λ6563 | N+B | ⋯ | ⋯ | ⋯ |
| [S II] λ6716/6731 | N+B | Yes | Fixed | ⋯ |

**Note.** For each line, N and B represent narrow (FWHM ⩽ 1200 km s$^{-1}$) and broad (1200 km s$^{-1}$ ⩽ FWHM ⩽ 10, 000 km s$^{-1}$) Gaussian components, respectively. The Fe II line complex model is adopted from Tsuzuki et al. (2006). The Center Spacing, FWHM, and Flux Ratio columns are described in Section 3.4. The line spacing of the [O III] lines is not fixed in the fitting process due to their high S/N.

[N II], we used the best-fit central wavelength of the stronger line in the doublet to constrain the weaker one with a fixed theoretical wavelength spacing and the same line width. In the fitting process, we fix the [O III] doublet ratio to 2.99 (e.g., Storey & Zeippen 2000; Dimitrijević et al. 2007) and the [N II] doublet ratio to 2.96 (e.g., Greene & Ho 2005; Wu et al. 2018; Jun et al. 2020). We are unable to measure the broad [N II] emission because it is degenerate with other line components, so we choose to remove the broad [N II] component from our analysis.

The Mg II emission line is modeled with a Fe II emission complex and a single Gaussian component (1G+1Fe) or a double-Gaussian (2G+1Fe) component. We adopted the Fe II template from Tsuzuki et al. (2006). The Fe II template from Vestergaard & Wilkes (2001) was also tested in our fitting process, but the results did not change significantly. Both the 1G+1Fe and 2G+1Fe models fit the Mg II line reasonably well, but an $F$-test yields a probability $p$-value of 0.54, implying that there is no significant improvement by adding an additional Gaussian component. We therefore adopt the 1G+1Fe profile to model Mg II line.

We did not correct for extinction based on the Balmer decrement (i.e., the ratio of Hα/Hβ) in this work because the contributions from the AGN and from star formation in the host galaxy cannot be easily disentangled. However, if we assume an intrinsic ratio of 2.86 for Hα/Hβ (e.g., Osterbrock 1989), the estimated $A_V \sim 4.6$ mag is lower than the value derived from the AGN component in the SED analysis ($A_V \sim 20$–60 mag; Eisenhardt et al. 2012; Assef et al. 2015, 2016). This implies that a significant fraction of the ionized hydrogen emission is from outflowing material and the host galaxy. Thus, in comparison with the AGN emission, the ionized hydrogen emission is overall experiencing less extinction.

## 4. Results and Discussion

### 4.1. FIR Emission and Dust Properties

W1904+4853 shows a rapidly rising SED around rest frame 4 μm as shown in Figure 2, followed by a gradual rise toward 100 μm, and a Rayleigh–Jeans falloff at FIR to submillimeter wavelengths. The IR radiation dominates the bolometric luminosity ($L_{\rm IR} = L_{2-1000\mu m} \sim 1.1 \times 10^{13} L_\odot$). Similar to that of other high-$z$ Hot DOGs (C. W. Tsai et al. 2023, in preparation), this SED shape suggests a continuous dust temperature distribution in the galaxy. The dust is heated to much hotter temperatures than in other IR-luminous galaxies. The high luminosity of W1904+4853 is due to the powerful, dust enshrouded AGN, with most energy released in the rest-frame MIR and FIR.

We applied a model with a single power-law (SPL) radial dust density distribution from Tsai et al. 2023, in preparation to fit the NIR to submillimeter SED. This SPL dust model assumes a continuous, spherically symmetric, optically thin dust cocoon in thermal equilibrium surrounding the central AGN. In addition to a scaling factor for matching the SED template with the observed fluxes, this model only relies on the representative hottest dust temperature ($T_{\rm h}$), representative coldest dust temperature ($T_{\rm c}$), and the power index $p$ of the dust grain number density, with an assumed dust emissivity $\beta = 1.5$. The fitting result is shown as the gray curve in Figure 2. The best-fit SPL dust SED model suggests $T_{\rm h} = 539 \pm 6$ K, $T_{\rm c} = 37 \pm 3$ K, and the power-law index of dust grain number density with radius $p = -0.82 \pm 0.02$. This model implies a total dust mass of $5^{+3}_{-2} \times 10^7 M_\odot$ for W1904+4853. These physical values are consistent with the range seen in high-$z$ Hot DOGs (Tsai et al. 2023, in preparation).

Utilizing the SPL model, our estimate indicates that the radius of the dust with a temperature of $T_{\rm h} = 539 \pm 6$ K encircling the core of W1904+4853 is roughly 20 pc d, implying that the light-travel time is on the order of 100 yr. While hard photons can quickly destroy dust grains after the dust absorption, this represents the lower limit on the timescale for the hot dust to persist after exposure to the central heating AGN. The equilibrium between dust destruction and replenishment can prolong this timescale significantly.

### 4.2. SMBH Mass and Eddington Ratio

With the assumption that the broad emission lines of the permitted lines are due to virialized motions of emitting gas around the central SMBH, the single-epoch mass of the BH $M_{\rm BH}$ can be estimated. As shown in Figure 1, the Hα and Hβ lines are both broad (FWHM >2000 km s$^{-1}$) and highly blueshifted ($\Delta v < -500$ km s$^{-1}$). These highly blueshifted Balmer lines could indicate that the emitting gas is outflowing (Jun et al. 2020), not suitable for $M_{\rm BH}$ estimate. In contrast, the Mg II line shows a systematic broad line, much less blueshifted, implying that Mg II light might originate around the SMBH and escape the dust cocoon through scattering (e.g., Assef et al. 2016, 2020). Outflows do not appear to significantly affect the Mg II line shape for W1904+4853. However, we cannot rule out the possibility that the broad Mg II line is associated with an outflow, rather than with the virialized gas around the SMBH. This caveat should be noted when considering the following discussions about the $M_{\rm BH}$ estimate based on the virialized Mg II gas assumption. Assuming an Hα/Mg II line ratio of 2.85 (Shen et al. 2011; Roig et al. 2014) and an equal line width, the corresponding broad Hα line would have a luminosity of $10^{42.2}$ erg s$^{-1}$. This component cannot be well separated from the outflowing broad-line component in the line profile fitting. Thus, we cannot conclusively rule out the presence of the emission from the central virialized gas near the central AGN in the entangled broad Hα line emission.

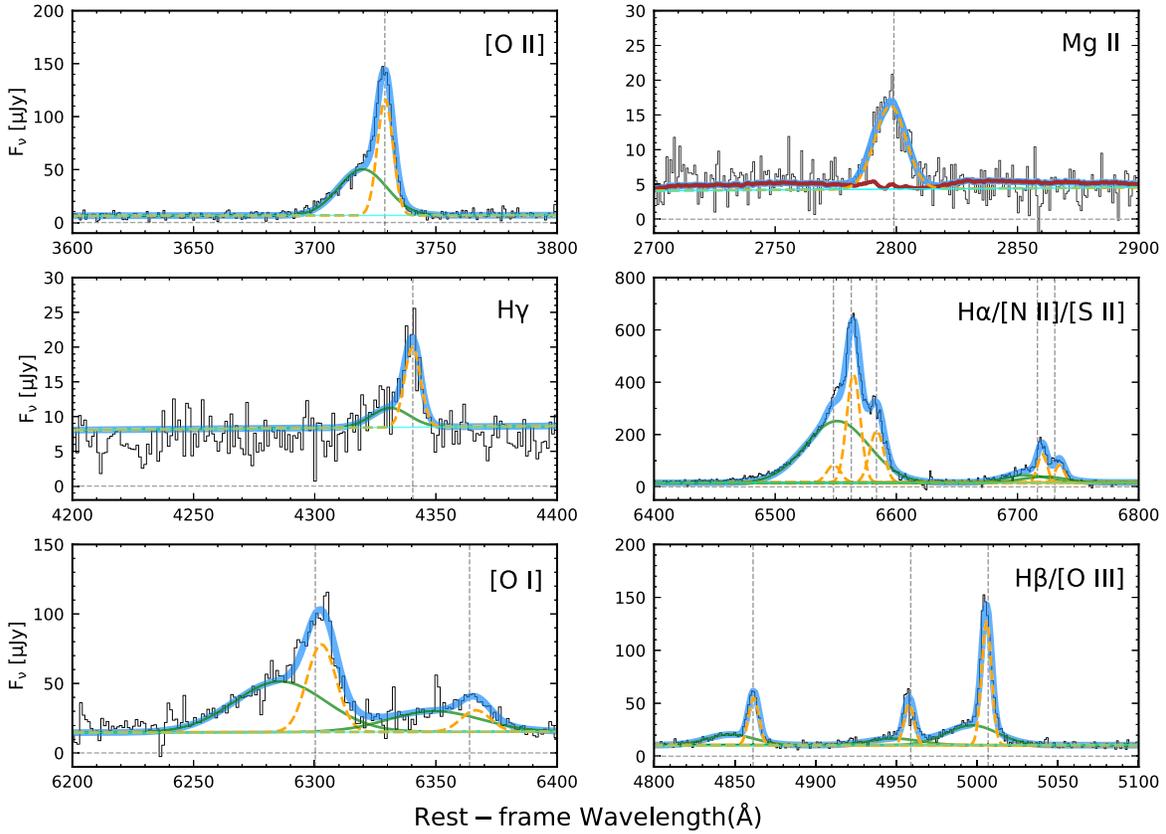

**Figure 4.** Observed spectrum of W1904+4853 around strong emission lines. The blue solid line shows the best-fit model in each panel. The green solid line and orange dashed lines represent the broad Gaussian and narrow Gaussian components, respectively. The cyan solid line represents the best-fit power-law continuum model ($F_\nu[\mu Jy] = (17.9 \pm 0.3) \times \lambda[\mu m]^{1.53 \pm 0.04}$). The brown solid line shows the Fe II line complex (Tsuzuki et al. 2006). We note that the apparent absorption in the O III 4959 Å emission line is an artifact due to a cosmic ray. This feature is not seen in the Keck/LRIS spectrum.

The Mg II line can be used to estimate BH masses (Wang et al. 2009; Tsai et al. 2018) and the calibration of Mg II determined SMBH masses with respect to those from H$\beta$ has been established (McLure & Dunlop 2004; Wang et al. 2009; Shen et al. 2011; Jun et al. 2015; Suh et al. 2015), assuming there is a close relationship between the BLR radius of the Mg II emitting region and the AGN continuum luminosity. In this paper, we adopt the Mg II-based SMBH mass formulation from Equation (10) of Wang et al. (2009):

$$\log\left(\frac{M_{BH}}{M_\odot}\right) = (7.13 \pm 0.27) + 0.5 \log\left(\frac{L_{3000}}{10^{44} \text{ erg s}^{-1}}\right)$$
$$+ (1.51 \pm 0.49) \log\left(\frac{FWHM_{Mg\,II}}{1000 \text{ km s}^{-1}}\right), \quad (1)$$

where $L_{3000}$ is the intrinsic (unobscured) monochromatic luminosity from the central AGN at rest-frame 3000 Å ($L_\lambda \equiv \lambda L_\lambda$), and FWHM$_{Mg\,II}$ is the full width at half maximum of the Mg II line profile.

The UV luminosity of W1904+4853 cannot be directly measured because of the high internal UV/optical extinction toward its central region. Instead, we estimate $L_{3000}$ using $L_{3000} \sim 0.19 \times L_{bol}$ based on the SED models for type 1 (broad-line) quasars from Richards et al. (2006). This implies $L_{3000} = 2.1 \times 10^{12} L_\odot$. With the FWHM of Mg II of $1600^{+200}_{-200}$ km s$^{-1}$, Equation (1) yields $\log(M_{BH}/M_\odot) = 8.4 \pm 0.1$ (stat.) $\pm 0.3$ (sys.). We note that the FWHM of Mg II in W1904+4853 is at the lower limit of measurements ($< 2500$ km s$^{-1}$) in Wang et al. (2009), with potential undersampling. To evaluate the reliability of Mg II line in estimating BH mass for AGNs with FWHM spanning from 800–2500 km s$^{-1}$, we utilized the AGN sample from Coffey et al. (2019) and followed the methodological approach mentioned in Wang et al. (2009). The result is consistent with the reported systematic uncertainty of 0.3 dex (Wang et al. 2009).

For W1904+4853, the Eddington ratio $\eta = L_{AGN}/L_{Edd} = 1.4^{+0.3}_{-0.2}$ (stat.)$^{+1.2}_{-0.7}$ (sys.) for $L_{bol} = 1.1 \times 10^{13} L_\odot$ and $\log(M_{BH}/M_\odot) = 8.4 \pm 0.4$, assuming the bolometric luminosity is mainly from the obscured AGN. Systematic uncertainties from the SMBH mass estimate (Equation (1)) dominate the uncertainty in $\eta$. The extremely high luminosity of high-redshift Hot DOGs is thought to be powered by embedded SMBHs with accretion rates close to the Eddington limit (Tsai et al. 2018; Wu et al. 2018). W1904+4853 has a similarly high accretion rate, potentially exceeding the Eddington limit.

### 4.3. Host Galaxy

The UV-NIR SED ($<1$ $\mu$m) of W1904+4853, as described in Section 3, is dominated by the host galaxy because the AGN within is highly obscured. We used empirical SED templates to decompose the UV-NIR SED into two components: a starburst population and a highly obscured starburst population. The fitting result shows that the starburst component dominates the UV-optical SED, while a young stellar population with extinction dominates the observed NIR ($J$, $H$, and $K$) bands.

Table 3
Emission Lines of W1904+4853

| Line | Flux (10$^{-16}$ erg cm$^{-2}$ s$^{-1}$) | log $L$ (erg s$^{-1}$) | log $L_{bl}$ (erg s$^{-1}$) | $\Delta v_{bl}$ (km s$^{-1}$) | FWHM$_{bl}$ (km s$^{-1}$) | EW$_{bl}$ (Å) | log $L_{nl}$ (erg s$^{-1}$) | $\Delta v_{nl}$ (km s$^{-1}$) | FWHM$_{nl}$ (km s$^{-1}$) | EW$_{nl}$ (Å) |
|---|---|---|---|---|---|---|---|---|---|---|
| Mg II | 9.5 ± 0.2 | $41.77^{+0.03}_{-0.04}$ | $41.77^{+0.03}_{-0.04}$ | $-160^{+60}_{-60}$ | $1600^{+200}_{-200}$ | $44^{+4}_{-3}$ | ... | ... | ... | ... |
| [O II] λ3727 | 36.0 ± 0.3 | $42.51^{+0.01}_{-0.01}$ | $42.25^{+0.01}_{-0.02}$ | $-580^{+30}_{-30}$ | $1880^{+50}_{-50}$ | $161^{+5}_{-5}$ | $42.16^{+0.02}_{-0.02}$ | $164^{+4}_{-4}$ | $610^{+10}_{-10}$ | $130^{+4}_{-4}$ |
| Hγ | 2.8 ± 0.2 | $41.24^{+0.06}_{-0.07}$ | $40.80^{+0.19}_{-0.27}$ | $-700^{+600}_{-300}$ | $1200^{+600}_{-500}$ | $6^{+4}_{-3}$ | $41.04^{+0.09}_{-0.13}$ | $-10^{+40}_{-50}$ | $500^{+100}_{-100}$ | $11^{+3}_{-3}$ |
| Hβ | 7.4 ± 0.2 | $41.82^{+0.02}_{-0.02}$ | $41.52^{+0.05}_{-0.06}$ | $-900^{+100}_{-100}$ | $2100^{+300}_{-300}$ | $36^{+4}_{-5}$ | $41.51^{+0.01}_{-0.01}$ | $25^{+9}_{-9}$ | $460^{+30}_{-30}$ | $35^{+3}_{-3}$ |
| [O III] λ4959 | 6.9 ± 0.2 | $41.53^{+0.02}_{-0.02}$ | $41.20^{+0.04}_{-0.04}$ | $-500^{+100}_{-100}$ | $2200^{+200}_{-200}$ | $24^{+3}_{-3}$ | $41.25^{+0.02}_{-0.02}$ | $-64^{+5}_{-5}$ | $430^{+20}_{-20}$ | $28^{+2}_{-2}$ |
| [O III] λ5007 | 15.5 ± 0.2 | $42.01^{+0.02}_{-0.02}$ | $41.68^{+0.04}_{-0.04}$ | $-500^{+100}_{-100}$ | $2200^{+200}_{-200}$ | $70^{+10}_{-10}$ | $41.74^{+0.02}_{-0.02}$ | $-63^{+5}_{-5}$ | $430^{+20}_{-20}$ | $83^{+6}_{-5}$ |
| [O I] λ6300 | 18.3 ± 0.3 | $42.15^{+0.01}_{-0.01}$ | $41.95^{+0.02}_{-0.02}$ | $-690^{+50}_{-50}$ | $2120^{+60}_{-60}$ | $117^{+5}_{-5}$ | $41.70^{+0.03}_{-0.03}$ | $122^{+9}_{-9}$ | $680^{+30}_{-30}$ | $65^{+4}_{-4}$ |
| [O I] λ6363 | 6.1 ± 0.2 | $41.68^{+0.01}_{-0.01}$ | $41.56^{+0.02}_{-0.02}$ | $-690^{+50}_{-50}$ | $2100^{+60}_{-60}$ | $47^{+2}_{-2}$ | $41.08^{+0.05}_{-0.05}$ | $121^{+9}_{-9}$ | $670^{+30}_{-30}$ | $16^{+2}_{-2}$ |
| [N II] λ6548 | ... | $41.67^{+0.01}_{-0.01}$ | ... | ... | ... | ... | $41.67^{+0.01}_{-0.01}$ | $7^{+2}_{-2}$ | $627^{+7}_{-7}$ | $62^{+1}_{-1}$ |
| Hα | ... | $43.00^{+0.01}_{-0.01}$ | $42.87^{+0.01}_{-0.01}$ | $-540^{+10}_{-10}$ | $2810^{+10}_{-10}$ | $981^{+7}_{-7}$ | $42.44^{+0.01}_{-0.01}$ | $83^{+1}_{-1}$ | $602^{+4}_{-4}$ | $368^{+4}_{-4}$ |
| [N II] λ6583 | ... | $42.13^{+0.01}_{-0.01}$ | ... | ... | ... | ... | $42.13^{+0.01}_{-0.01}$ | $7^{+2}_{-2}$ | $624^{+7}_{-7}$ | $182^{+3}_{-3}$ |
| [S II] λ6716 | ... | $42.02^{+0.04}_{-0.02}$ | $41.76^{+0.07}_{-0.03}$ | $-420^{+60}_{-40}$ | $1900^{+50}_{-50}$ | $80^{+10}_{-10}$ | $41.68^{+0.04}_{-0.04}$ | $177^{+3}_{-3}$ | $408^{+9}_{-9}$ | $65^{+2}_{-2}$ |
| [S II] λ6731 | ... | $41.87^{+0.02}_{-0.06}$ | $41.66^{+0.04}_{-0.11}$ | $-420^{+60}_{-40}$ | $1900^{+50}_{-50}$ | $60^{+10}_{-10}$ | $41.46^{+0.02}_{-0.02}$ | $177^{+3}_{-3}$ | $408^{+9}_{-9}$ | $40^{+2}_{-2}$ |

**Note.** The fluxes were calculated by fitting one/two Gaussian profiles to the emission-line profiles, measuring the integrated flux under the profile. Mixing with other lines, the integrated flux of [N II], Hα, and [S II] cannot be estimated individually. $L$ represents the emission-line luminosity derived from best-fit model and $\Delta v$ represents the offset of the Gaussian center for each broad/narrow component with respect to the systemic redshift of $z = 0.415$ determined by multiple lines following the scheme of Eisenhardt et al. (2023, in preparation). Subscripts "bl" and "nl" indicate broad line and narrow line, respectively.

To further extract the properties of the host galaxies from the SED and optical spectrum, the following subsections discuss how the star formation rate (SFR) and stellar mass are estimated. The SFR is derived using the ultraviolet continuum emission (Section 4.3.1) and narrow component of [O II] line emission (Section 4.3.2). We estimated the stellar mass according to the model from Bell et al. (2003) in Section 4.3.3.

*4.3.1. UV-based SFR*

The unobscured SFR can be estimated using the SED of the host galaxy. A common practice is combining UV and IR measurements and estimating the dust-corrected SFR using the calibrated relation that relates UV and IR luminosities to the SFR (Kennicutt & Evans 2012). In the case of W1904+4853, our best-fit model implies that dust fully obscures the AGN at observed optical wavelengths except for the broad emission lines leaking out of the dust cocoon around the nucleus. Under the assumption that the UV continuum is solely due to galaxy light with no AGN emission leaking through, the UV-optical SED is dominated by a young stellar population according to our SED modeling (Section 3.2), which would imply an SFR of W1904+4853's host galaxy. This estimated SFR$_{UV}$ from UV luminosity can be considered as a lower limit.

We adopted the models from Assef et al. (2010) to fit the stellar components for the host galaxy in Section 3 and use the models of Bruzual & Charlot (2003) to estimate the host galaxy's stellar population age and extinction here.

Assuming no UV extinction from the host galaxy toward the young star population of W1904+4853, the average $L_\nu^{UV}$ = 4.4 × 10$^{28}$ erg s$^{-1}$ Hz$^{-1}$ is calculated by integrating over the best-fit SED at 1500–2800 Å, implying a SFR$_{UV}$ ∼ 6 $M_\odot$ yr$^{-1}$ for a Salpeter initial mass function (IMF) of a stellar population with stars of 0.1–100 $M_\odot$ (Kennicutt 1998, Equation (1)). This method applies to a galaxy with continuous star formation over a timescale of 10$^8$ yr or longer. The estimated SFR would be significantly higher if the host galaxy had a shorter, more intensive star formation episode. Therefore, the derived SFR$_{UV}$ ∼ 6 $M_\odot$ yr$^{-1}$ can be considered as the lower limit. Moreover, the SFR$_{UV}$ of the host galaxy for W1904+4853 would be higher if intrinsic host extinction were included.

To estimate the impact of extinction on the host galaxy SFR estimate, we apply dust reddening to the galaxy SED models. We used the SED of a simple stellar population (SSP) with solar metallicity and a Salpeter IMF from Bruzual & Charlot (2003) and the extinction curve from Fitzpatrick & Massa (2007) to fit the continuum emission of W1904+4853 over UV-to-optical wavelengths. The photometric data are corrected for Milky Way extinction and AGN emission-line contamination prior to the modeling. We note that due to possible contamination from C IV and C III] lines discussed in Section 3.1, we adopted the predicted values of 4.12 mJy for NUV based on our best-fit SED.

The best-fit SSP model from Bruzual & Charlot (2003) implies a stellar population with an age of 0.02 Gyr and a V-band extinction of $A_V = 0.9^{+0.2}_{-0.1}$ mag. The result implies that the young stellar population from a recent starburst episode dominates the optical continuum emission from the host galaxy. We then reestimated the average extinction-corrected $L_\nu^{UV}$ of the host galaxy, finding SFR$_{UV} = 52^{+32}_{-20}$ $M_\odot$ yr$^{-1}$. The uncertainty is mainly due to statistical uncertainties in the SED fitting. We note that this estimate is based on a model with a decaying star formation history. If a constant SFR is adopted, the best-fit extinction is higher by 0.5 mag and the extinction-corrected SFR$_{UV}$ is higher by ∼60%. We also examined using the Chabrier IMF, but there is no significant change in the results.

The IR luminosity driven by star formation activity contributes just ∼3% of the bolometric luminosity of W1904+4853 and ∼8% of the FIR luminosity assuming the FIR–SFR relation of Kennicutt (1998). Therefore, the energy output of this system is dominated by the obscured AGN, as discussed in Section 4.1.

### 4.3.2. [O II]-based SFR

For moderately redshifted systems, optical observation of the [O II]–3727 Å doublets (3726 Å and 3729 Å) provide an additional SFR tracer (Kennicutt 1998), although it is often affected by significant dust attenuation and suffers from large systematic uncertainties (i.e., Moustakas et al. 2006, and references therein). Using metallicities derived from the [N II] λ6584/[O II]λ3727 ratio, Zhuang & Ho (2019) refine a metallicity-dependent SFR estimator based on [O II]. They applied this method to estimate the SFRs of a sample of ∼5500 type 2 AGNs, and the results are consistent with independent SFR obtained from the stellar continuum of the host galaxies. By removing the blueshifted component using the double-Gaussian decomposition of the observed [O II] profile, the narrow [O II] component provides an independent SFR tracer for W1904+4853.

We assume that the narrow [O III] emission is driven by the AGN activity and the narrow [O II] emission mainly arises from the star-forming region in the host galaxy. According to Zhuang & Ho (2019, Equation (7)), the calibrated [O II] SFR for AGNs is

$$\text{SFR}_{\text{[OII]}}(M_\odot \text{ yr}^{-1}) = \frac{5.3 \times 10^{42}(L_{\text{[O II]}} - 0.109\, L_{\text{[O III]}})}{(-4373.14 + 1463.92x - 163.045x^2 + 6.04285x^3)}, \quad (2)$$

where $L_{\text{[O II]}}$ and $L_{\text{[O III]}}$ are the total [O II] and narrow [O III] luminosities in erg s$^{-1}$, respectively. These numbers for W1904+4853 are corrected for the extinction from dust in the Milky Way and host galaxy ($A_V = 0.9^{+0.2}_{-0.1}$ mag) as discussed in Section 4.3.1. The variable $x$ in Equation (2) is the oxygen abundance $x \equiv \log(\text{O/H}) + 12$. We adopt an oxygen abundance of 8.64, which is the average value of that in Seyfert 2 AGNs (Dors et al. 2020).

We note that W1904+4853 is significantly brighter than most of the galaxies in the Zhuang & Ho (2019) sample. Using the extinction-corrected emission-line luminosities, we estimate an SFR$_{\text{[O II]}}$ of ∼45 $M_\odot$ yr$^{-1}$ for the host galaxy of W1904+4853. However, if a fraction of the [O II] is shock ionized (Maddox 2018), the SFR$_{\text{[O II]}}$ is an upper limit. We also note that the observed 610 km s$^{-1}$ line width of narrow [O II] line emission is higher than the typical value in luminous starbursts (∼<= 300 km s$^{-1}$; Banerji et al. 2011). This system could be at a late stage of dynamically settling in galaxy-galaxy merger event, as in the cases of ULIRGs (∼ 900 km s$^{-1}$; Chen et al. 2019).

### 4.3.3. Stellar Mass

At low redshift, the stellar mass of the host galaxy is usually dominated by the old and low-mass stellar population, which produces the majority of the rest-frame NIR emission of the system. Here we first estimate the mass-to-light ($M/L$) ratio of the W1904+4853 host galaxy, and derive the stellar mass using the rest-frame $K$-band flux from the best-fitting results discussed in Section 3.

Using the correlations between $ugriz$ colors and $M/L$ ratios reported by Bell et al. (2003), we estimated the stellar $M/L$ ratio for the host galaxy in the W1904+4853 system. Stellar $M/L$ ratios are given by $\log_{10}(M/L_K) = -0.209 + (0.197 \times (g - r))$, where the $M/L$ ratio is in solar units. To obtain the rest-frame $g$-, $r$- and $K$-band fluxes, we use the best-fit SED model from Figure 2, which yields a K-corrected value of $(g - r) = 0.1$, and a correction factor $K = 0.50$. This implies $M/L_K = 0.65$ and $L_K = 4.6 \times 10^{10} L_\odot$. The typical $M/L$ ratio uncertainty is ∼0.1 dex and is dominated by systematic uncertainties. The stellar mass of the host galaxy of W1904+4853 is then estimated to be $3 \times 10^{10} M_\odot$, which follows the local relation between $M_{\text{BH}}$ and host galaxy stellar mass (Treu et al. 2007; Bennert et al. 2010, 2011; Assef et al. 2015). The stellar mass varies by less than 50% if the different derived $M/L$ values from various combinations of $griz$—colors and NIR photometry are adopted.

The NIR SED ($J$, $H$, and $K$ bands) is dominated by a highly obscured starburst population with $A_V \sim 7.5$ mag, as shown in Figure 2. We note that the implied stellar mass from the highly obscured Im component could reach $1.1 \times 10^{11} M_\odot$ if $M_{\text{star}}/L_K = 1$ is adopted. Some contribution from AGN emission to the $K$-band flux would imply that we are overestimating the extinction-free Im component of the host galaxy. The estimated $1.1 \times 10^{11} M_\odot$ stellar mass is therefore an upper limit.

### 4.4. High SFR and Strong AGN activity in W1904+4853

#### 4.4.1. Spectral Classification

The rest-frame optical SED of W1904+4853 shows little contribution from the powerful AGN although it dominates the IR emission. This suggests that the observed narrow emission lines potentially arise from ionized gas from the host galaxy. To distinguish the dominating photoionizing processes, we evaluate the emission-line ratios of W1904+4853 using standard BPT diagrams (e.g., Baldwin et al. 1981). As shown in Figure 5, we use three diagnostics which incorporate the ratios of [N II]/H$\alpha$, [S II]/H$\alpha$, [O I]/H$\alpha$, and [O III]/H$\beta$ (Kewley et al. 2001; Kauffmann et al. 2003; Kewley et al. 2006). The locations on the BPT diagrams are similar to those for the $z = 0.83$ Hot DOG J012611.96052909.6 (Jun et al. 2020) yet closer to the starburst region compared to the more luminous and distant Hot DOGs in Jun et al. (2020) with AGN dominance. These similarities indicate that W1904+4853 is rather complicated. The properties of W1904+4853 are highly consistent with that this system, like some red quasars, hosts powerful, obscured AGN and also strong star formation activity. For these red quasar systems, the high SFR may be enhanced by early galaxy interactions, making them located in the composite area of the BPT diagram (e.g., Ellison et al. 2016a, 2016b).

#### 4.4.2. Comparison with Star-forming Main Sequence

The main sequence (MS) of star formation in galaxies shows the correlation between SFR and stellar mass for star-forming galaxies (e.g., Brinchmann et al. 2004; Daddi et al. 2007; Elbaz et al. 2007; Noeske et al. 2007), which changes with redshift (e.g., Pannella et al. 2009; Whitaker et al. 2012; Moustakas et al. 2013; Schreiber et al. 2015). To compare W1904+4853 at $z = 0.415$ with the MS, we considered galaxies with redshift between 0.4 and 0.5 from the PRIsm MUlti-Object Survey (PRIMUS; Coil et al. 2011; Moustakas et al. 2013). We re-derived their SFRs and stellar masses using the same methods applied to W1904+4853 in Section 4.3 to avoid the systematic biases between SFR and stellar mass estimators. The result is shown in Figure 6. The large magenta point shows W1904

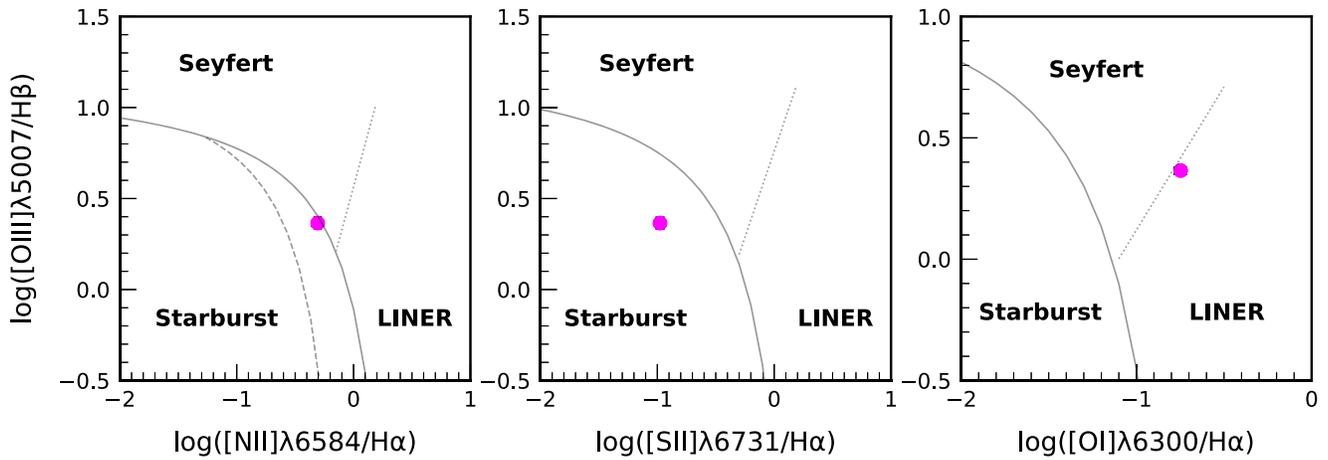

**Figure 5.** BPT diagrams for W1904+4853. The magenta points denote the line ratios derived from the narrow-line fluxes of W1904+4853. The error bars are smaller than the symbols; thus, they are not visible in this plot. Solid lines show the demarcation between starburst galaxies and AGN defined by Kewley et al. (2001) for all panels. The dashed line denotes the dividing line between star-forming and composite star-forming (Kauffmann et al. 2003). The dotted line separates Seyfert and LINER-type AGNs (Kewley et al. 2006).

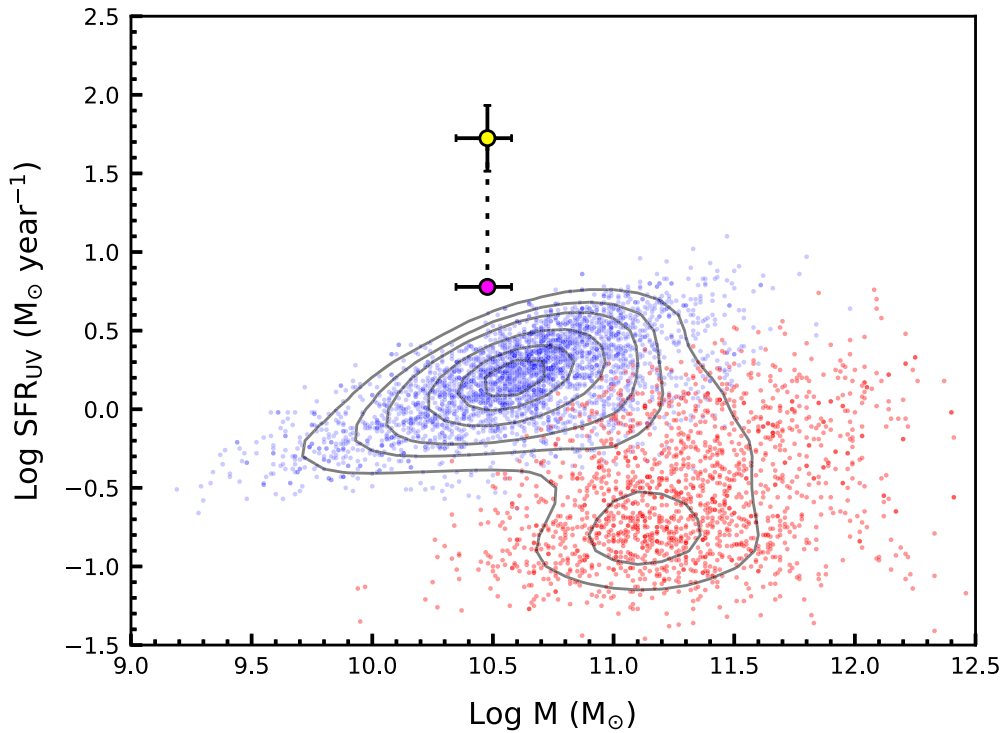

**Figure 6.** SFR vs. stellar mass of W1904+4853 compared with galaxies in the PRIMUS sample (Moustakas et al. 2013). The magenta point represents the lower limit to the SFR based on the UV continuum without considering the extinction from the host galaxy. The yellow point represents the extinction-corrected SFR from the host galaxy, which is close to the SFR inferred from the narrow [O II] line. Both SFRs and stellar masses in the PRIMUS sample are recalculated using the methods discussed in Sections 4.3.1 and 4.3.3. The blue dots represent galaxies with luminosity dominated by the young stellar population in the SED fitting, while the red dots represent galaxies dominated by the old stellar population.

+4853 with the lower limit of SFR calculated from the UV continuum of the SED model, while the small blue and red points are PRIMUS galaxies with SFRs. The large yellow point shows W1904+4853 with its $SFR_{UV}$ derived using the BC03 model on the extinction-corrected host galaxy SED. We note that the density of quiescent galaxies in Figure 5 is much lower than reported in Moustakas et al. (2013) because using UV continuum to estimate the SFRs is unreliable for quiescent galaxies with depressed star formation. Those systems are omitted in the figure.

As discussed in previous sections, the SED of W1904+4853's host galaxy is dominated by a starburst component. Considering extinction from the host galaxy, the SFRs estimated from the UV continuum and [O II] line are both $\gtrsim 45\,M_\odot$ yr$^{-1}$, significantly higher than the MS as shown in Figure 5. This result indicates that the activity of the obscured AGN in W1904+4853 has not yet significantly impacted the star formation activity of the host galaxy. The SMBH in W1904+4853, similar to high-$z$ Hot DOGs (Eisenhardt et al. 2012; Wu et al. 2012; Tsai et al. 2018; Wu et al. 2018), has likely just been activated recently and is accreting in an extremely dusty environment while the whole galaxy is still experiencing an ongoing starburst episode.

### 4.4.3. Ionized Gas Outflows

Figure 4 shows the profiles of the [O II], [O III], H$\beta$, H$\gamma$, and H$\alpha$ lines, with their broad and narrow components highlighted. Similar to many type 2 quasars (e.g., Zakamska & Greene 2014; Kang et al. 2017), the [O II], [O III] doublets, and hydrogen Balmer lines of W1904+4853 are asymmetric and blueshifted. These spectroscopic features could be indications of ionized gas outflows from their central obscured AGNs. This type of outflow is often modeled by biconical motions where the blue wing is less affected by obscuration than the red, leading to an asymmetric line profile (e.g., Crenshaw et al. 2010; Zakamska & Greene 2014; Bae & Woo 2016; Jun et al. 2020). This model implies that these asymmetric and blueshifted emission lines do not come from the vicinity of the BH. We note that the Mg II line profile used for $M_{BH}$ estimation (Section 4.2) does not show the same asymmetric and blueshifted line profile as the Balmer lines.

W1904+4853 is not unique with its ionized gas outflows among hyperluminous obscured AGNs. The spectra of almost all high-$z$ Hot DOGs observed with Very Large Telescope/XSHOOTER or Keck/NIRSPEC (Finnerty et al. 2020; Jun et al. 2020) show broad and blueshifted [O III] profiles. In previous work, the kinematic quantities of the outflowing ionized gas such as mass outflow rate ($\dot{M}_{out}$), energy injection rate ($\dot{E}_{out}$), and momentum flux ($\dot{P}_{out}$), are estimated using the broad [O III] profile under the assumption of a uniform, filled spherical/biconical outflow geometry (e.g., Maiolino et al. 2012). Those high-$z$ Hot DOGs host powerful AGN and drive a strong outflow comparable to the most powerful outflow systems in the literature (Jun et al. 2020). These kinematic properties scale with the total [O III] luminosity $L_{[OIII]}$.

Using the observed [O III] profile properties listed in Table 3 and the recipes described in Jun et al. (2020, Equations (1) and (2)), we estimate that the outflows in W1904+4853 contain $\log(M_{gas}/M_\odot) = 5.95 \pm 0.01$, $\dot{M}_{out} = 1.66 \pm 0.04\ M_\odot\ \mathrm{yr}^{-1}$, $\log \dot{E}_{out}\ (\mathrm{erg\ s}^{-1}) = 42.25 \pm 0.02$, and $\log \dot{P}_{out}\ (\mathrm{dyn}) = 34.29 \pm 0.01$. Although the bolometric luminosity of W1904+4853 is comparable to the sample in Jun et al. (2020), its $L_{[OIII]}$ is 2 orders of magnitudes lower than those high-$z$ Hot DOGs. With the consideration of the lower [O III] line luminosity, W1904+4853 still follows the same correlations between the kinematic properties of the [O III] outflow and $L_{[OIII]}$ among the powerful outflow systems (see Figure 7 in Jun et al. 2020).

The lower [O III] luminosity implies a smaller mass outflow rate for the W1904+4853 system, significantly smaller than the SFR estimated from the UV-optical SED in Section 4.3.1. The outflows in W1904+4853 do not appear to be strong enough yet to deplete the interstellar gas in the galaxy and suppress ongoing star formation.

### 4.5. W1904+4853 as a Low-z Example of a Transitional AGN

The analysis of Section 4.2 implies that W1904+4853 hosts an SMBH that is accreting at close to the Eddington limit. Although the powerful, yet heavily obscured AGN in W1904+4853 drives a detectable outflow, the SFR is higher than in MS galaxies of similar mass and redshift, making the host galaxy a starburst system. The central AGN in W1904+4853 has not significantly hindered the host galaxy's star formation. The rapid AGN accretion hidden within the dust cocoon in the actively star-forming galaxy indicates that W1904+4853 is indeed a system in the transitional stage of galaxy evolution.

Assuming the number density of Hot DOGs is comparable to that of the unobscured QSOs with similar luminosities (Assef et al. 2015), the number of low-$z$ Hot DOGs can be estimated from QSO surveys if the same relation still holds for the low-$z$ systems. In SDSS DR7,[12] there are nine type 1 QSOs at low redshift ($z < 0.5$) with $L_{bol} \geqslant 10^{13}\ L_\odot$ from 11,663 deg$^2$ at $z \sim 0.365$–0.465 (Abazajian et al. 2009). The comoving volume within $z \sim 0.365$–0.465 is 11.121 Gpc$^3$ (Wright 2006), indicating the volume density of type 1 QSOs is $\sim 3$ Gpc$^{-3}$ in this redshift range. Therefore, we expect $\sim 30$ hyperluminous QSOs/Hot DOGs at $z \sim 0.365$–0.465 for the whole sky based on these assumptions. W1904+4853 could yield a good model to inspire a comprehensive search for low-$z$ obscured AGN systems such as low-$z$ Hot DOGs.

## 5. Conclusions

We report the highly dust-obscured quasar WISE J190445.04+485308.9, or W1904+4853, located in the relatively low-$z$ universe ($z = 0.415$). Its UV-to-FIR SED shows a broad, MIR plateau and a falloff at the FIR-to-submillimeter wavelengths. This SED shape indicates that the emission of this system is dominated by hot dust emission. The dominant emission from hot dust in this system can be explained by an accreting BH hidden in its dusty cocoon, where the dust reprocesses the primary energy released from the accretion disk. From spectroscopic and SED analyses, we find the properties of W1904+4853 are as follows:

1. The broad Mg II line is symmetric and only marginally blueshifted, suggesting that the line originates around the SMBH, and that the light escapes from the dust cocoon by scattering. This assumption could be tested by future spectropolarimetric observations. The broad Mg II line implies a BH mass $\log(M_{BH}/M_\odot) = 8.4 \pm 0.4$. Combined with the well-determined bolometric luminosity $L_{bol} = 1.1 \times 10^{13}\ L_\odot$, the Eddington ratio is $\eta_{Edd} \sim 1.4^{+0.3}_{-0.2}(\mathrm{stat.})^{+1.2}_{-0.7}(\mathrm{sys.})$, indicating that the SMBH is accreting at close to the Eddington limit.

2. IR dust emission ($L_{IR}$) contributes $\sim 99\%$ of the bolometric luminosity of W1904+4853, dominating the total energy output of this system. A power-law dust model suggests a dust temperature range between $539 \pm 6$ K and $37 \pm 3$ K. The dust close to the AGN is heated by the accreting SMBH to higher temperatures than of other IR-luminous galaxies (30–40 K; Melbourne et al. 2011; Magnelli et al. 2012). If the dust is in radiative equilibrium, the temperature range implies the dust has spread to a large radius, reaching kiloparsec scales.

3. The SED of the host galaxy is surprisingly dominated by the young stellar population, with an SFR of $\sim 6$–84 $M_\odot\ \mathrm{yr}^{-1}$. An independent estimate yields an SFR of $\sim 45\ M_\odot\ \mathrm{yr}^{-1}$ using the [O II] line. This star formation contributes only $\sim 8\%$ of the FIR luminosity, supporting that the IR dust emission is powered by central AGN.

---

[12] https://classic.sdss.org/dr7

4. Most of the optical emission lines observed in W1904+4853, including [O II], [O III], H $\gamma$, H$\beta$, and H$\alpha$, are significantly asymmetric and blueshifted. These line profiles are likely affected by outflowing ionized gas from the AGN.
5. The outflow mass is an order of magnitude less than the ongoing SFR of the host galaxy, suggesting that this system is at the early stage of the *blowout* phase when the outflow power has not yet fully launched to drive the star-forming material out of the system. The obscured AGN activity has not yet significantly affected the star formation activity in the host galaxy.

These properties of W1904+4853 show that it is in a transitional phase of galaxy evolution when the vigorous AGN activity has just turned on but has not yet impacted the star formation activity in the host galaxy significantly. It is highly similar to the conditions of the high-redshift Hot DOGs. W1904+4853 is the first confirmed low-redshift Hot DOG, making it a great local laboratory to study a heavily obscured AGN during its transitional stage in the host-SMBH coevolution. Deeper spectroscopic observation with higher sensitivity could reveal physical properties of the stellar population in its host galaxy, such as the 4000 Å break. Spatially resolved IFU observations of W1904+4853 using ground-based telescopes with adaptive optics, JWST, and CSST can potentially capture the AGN and starburst activities coexisting within this galaxy. Deep radio observations of W1904+4853 could assess the potential effect of radio mode feedback. More detailed investigations of W1904+4853 could help provide a more complete picture of astrophysical processes in Hot DOGs. The multiwavelength investigation of W1904+4853 discussed in this paper provides a foundation for future studies on high-resolution facilities such as using JWST, CSST, and VLBA to reveal the critical and detailed information on cold dust content, AGN energetics, and jet feedback of this obscured quasar in its transformation stage.

## Acknowledgments

The authors would like to thank A. J. Barth for valuable comments on this work. This work was supported by a grant from the National Natural Science Foundation of China (Nos. 11973051, 11673029). This work is also supported by the International Partnership Program of the Chinese Academy of Sciences, program No. 114A11KYSB20210010. Portions of this research were carried out at the Jet Propulsion Laboratory (JPL)/California Institute of Technology (Caltech) under a contract with the National Aeronautics and Space Administration (NASA).

R.J.A. was supported by FONDECYT grant Nos. 1191124 and 1231718, and by ANID BASAL project FB210003. H.D.J. was supported by the National Research Foundation of Korea (NRF) funded by the Ministry of Science and ICT (MSIT) (Nos. 2020R1A2C3011091, 2021M3F7A1084525, 2022R1C1 C2013543). S.E.L. was supported by a grant from the Chinese Academy of Sciences Presidents International Fellowship Initiative (No. 2019PM0017) and NSFC grant No. U1931110.

Some of the data reported here were obtained, and operations were enabled through the cooperation of the East Asian Observatory. The James Clerk Maxwell Telescope is operated by the East Asian Observatory on behalf of The National Astronomical Observatory of Japan; Academia Sinica Institute of Astronomy and Astrophysics; the Korea Astronomy and Space Science Institute; the National Astronomical Research Institute of Thailand; Center for Astronomical Mega-Science (as well as the National Key R&D Program of China with No. 2017YFA0402700). Additional funding support is provided by the Science and Technology Facilities Council of the United Kingdom and participating universities and organizations in the United Kingdom and Canada. UKIRT is owned by the University of Hawaii (UH) and operated by the UH Institute for Astronomy.

This publication makes use of data products from the Wide-field Infrared Survey Explorer, which is a joint project of the University of California, Los Angeles, and the Jet Propulsion Laboratory/California Institute of Technology, and NEO-WISE, which is a project of the Jet Propulsion Laboratory/California Institute of Technology. This research has made use of the NASA/IPAC Infrared Science Archive, which is funded by the National Aeronautics and Space Administration and operated by the California Institute of Technology. These infrared data products can be accessed via DOI:10.26131/IRSA1 for the AllWISE, 10.26131/IRSA180 for the Akari and DOI:10.26131/IRSA11 for the IRAS.

Some of the data presented in this paper were obtained from the Mikulski Archive for Space Telescopes (MAST) at the Space Telescope Science Institute. The specific observations analyzed can be accessed via DOI:10.17909/ths3-nr82 for the GALEX. NUV data, the Pan-STARRS1 DR2 Catalog at DOI:10.17909/s0zg-jx37 and the Spitzer Kepler Survey at DOI:10.17909/mkcw-8w83. The Pan-STARRS1 Surveys (PS1) and the PS1 public science archive have been made possible through contributions by the Institute for Astronomy, the University of Hawaii, the Pan-STARRS Project Office, the Max-Planck Society and its participating institutes, the Max Planck Institute for Astronomy, Heidelberg and the Max Planck Institute for Extraterrestrial Physics, Garching, the Johns Hopkins University, Durham University, the University of Edinburgh, the Queen's University Belfast, the Harvard-Smithsonian Center for Astrophysics, the Las Cumbres Observatory Global Telescope Network Incorporated, the National Central University of Taiwan, the Space Telescope Science Institute, the National Aeronautics and Space Administration under Grant No. NNX08AR22G issued through the Planetary Science Division of the NASA Science Mission Directorate, the National Science Foundation Grant No. AST-1238877, the University of Maryland, Eötvös Loránd University (ELTE), the Los Alamos National Laboratory, and the Gordon and Betty Moore Foundation.

*Facilities:* IRSA, Hale, Keck: I (LRIS), UKIRT, WISE, IRAS, JCMT.

## ORCID iDs

Guodong Li https://orcid.org/0000-0003-4007-5771
Chao-Wei Tsai https://orcid.org/0000-0002-9390-9672
Daniel Stern https://orcid.org/0000-0003-2686-9241
Jingwen Wu https://orcid.org/0000-0001-7808-3756
Roberto J. Assef https://orcid.org/0000-0002-9508-3667
Andrew W. Blain https://orcid.org/0000-0001-7489-5167
Tanio Díaz-Santos https://orcid.org/0000-0003-0699-6083
Thomas H. Jarrett https://orcid.org/0000-0002-4939-734X
Hyunsung D. Jun https://orcid.org/0000-0003-1470-5901
Sean E. Lake https://orcid.org/0000-0002-4528-7637
M. Lynne Saade https://orcid.org/0000-0001-7163-7015